# The effect of ozone oxidation on single-walled carbon nanotubes

J.M. Simmons[1], B.M. Nichols[2], S.E. Baker[2], Matthew S. Marcus[1,2], O.M. Castellini[1], C.-S. Lee[2], R.J. Hamers[2], M.A. Eriksson[1*]

[1]Department of Physics, University of Wisconsin - Madison

[2]Department of Chemistry, University of Wisconsin – Madison

Exposing single-walled carbon nanotubes to room temperature UV-generated ozone leads to an irreversible increase in their electrical resistance. We demonstrate that the increased resistance is due to ozone oxidation on the sidewalls of the nanotubes rather than at the end caps. Raman and x-ray photoelectron spectroscopy show an increase in the defect density due to the oxidation of the nanotubes. Using ultraviolet photoelectron spectroscopy we show that these defects represent the removal of π-conjugated electron states near the Fermi level, leading to the observed increase in electrical resistance. Oxidation of carbon nanotubes is an important first step in many chemical functionalization processes. Since the oxidation rate is controllable with short exposures, UV-generated ozone offers the potential for use as a low-thermal budget processing tool.

*e-mail: maeriksson@wisc.edu

**Introduction**

Nanotubes have been shown to be effective, though non-specific, sensors for low concentrations of gases such as $NO_2$, $NH_3$, and ozone, as well as several proteins.[1-6] The sensors operate by detecting a change in the electrical resistance due to adsorption of the analyte. Possible mechanisms for resistance change upon exposure to chemicals include charge transfer doping to the nanotube,[1,7] the modification of the potential barrier between the nanotube and metal electrodes,[8] or a direct change in the conduction channel (e.g., in the density of states) in the nanotube itself. Determining the relevant mechanism can be often difficult because several mechanisms can act simultaneously. For example, experiments indicate that the dominant resistance-changing effect of oxygen adsorption on nanotubes is not doping, but rather it is a change in work function of adjoining metal electrodes and a modification of the metal-nanotube Schottky barrier.[8,9] Nonetheless, theoretical work shows a sizeable charge transfer from chemisorbed oxygen to nanotubes.[10-12] These two results are not necessarily in conflict. Studies of the oxygen adsorption process indicate that oxygen adsorbs with very low coverages,[13] which would result in a low concentration of dopant carriers. Thus, both Schottky barrier height changes and doping may occur, with resistance changes dominated by the former.

Oxygen-nanotube interactions are particularly interesting because oxidation is an important first step in a number of chemical functionalization strategies.[14-17] One potentially useful form of oxygen is ozone, easily generated by ultraviolet excitation of molecular oxygen. The UV photochemistry of oxygen and carbon is quite complex, though UV-generated ozone has been demonstrated to be useful for the oxidation of carbonaceous contaminants on a number of substrates.[18-20] In brief, 185 nm UV light leads to the generation of ozone, and 254 nm UV light can dissociate the ozone molecule to atomic oxygen.[20,21] In addition, the rate of oxidation is increased when the carbon is exposed to both UV light and ozone, though the cause for the enhancement is not clear. The differing oxidation rates could be due to either a higher concentration of ozone near the carbon when the UV light illuminates it,[19] or to optically excited states in the carbon.[20] Though both ozone and atomic oxygen are highly reactive species, the formation of ozone is the initial step in photochemical oxidations.

In the case of fullerenes and nanotubes, ozone has been predicted to chemisorb via a [2+3] cycloaddition (Criegee's mechanism) creating a short-lived ozonide species that spontaneously decomposes into an epoxide or carbonyl.[22-26] In the case of nanotubes, Criegee's mechanism has been predicted to occur even in the absence of a pre-existing defect site.[26] Dissociation of ozone by 254 nm UV light can generate highly reactive singlet oxygen atoms and molecules,[21] which could either oxidize defects or create new defects. Indeed, density functional theory calculations have shown that chemisorption of singlet oxygen has a lower energy barrier than that of triplet oxygen.[27] Other methods of oxidation, such as acid reflux and air oxidation,[28-30] require either long process times or elevated temperatures and are primarily bulk processes, whereas ozone oxidation has been shown to oxidize nanotubes at room temperature and can be performed on individual nanotubes on a surface.[7,31-39] Ozone oxidation[40] thus offers the possibility of controllably oxidizing nanotubes that have been integrated into functional devices without affecting the device thermal budget. The feasibility of low temperature reactions with integrated nanotubes has been demonstrated recently.[41] Since extensive exposure to ozone will etch nanotubes and significantly degrade their properties, controlling the exposure conditions and understanding the consequences of the ozone oxidation is important.[34,37,42]

Here we measure the effect of UV-generated ozone on the physical and electronic structure of single-walled carbon nanotubes. Upon exposure to ozone, the resistance of the nanotubes increases, and, importantly, the increase is not reversible in our experiments, in contrast to previous reports.[43] In order to probe the cause for the resistance change, we use Raman spectroscopy, x-ray photoelectron spectroscopy (XPS) and ultraviolet photoelectron spectroscopy (UPS) to measure the changes in the chemical and electronic structure of the nanotubes. Due to their differing sensitivities and properties probed, it is only through the use of several techniques that we are able to obtain a comprehensive understanding of the effects of ozone exposure to carbon nanotubes. As discussed above for the case of molecular oxygen adsorption, transport measurements can be sensitive to a large number of variables, and thus require the use of other techniques to understand the chemistry involved. Raman spectroscopy allows us to correlate a resistance change with a change in defect density, and photoelectron spectroscopy ties the defects to the specific chemical species present. Though ozone will oxidize both the nanotube end caps and sidewalls, we show that the resistance increase in metal-nanotube-metal devices is not due to the oxidation of the nanotube end caps, or to a contact effect, but rather is due to the oxidation of the nanotube sidewalls. The oxidation leads to a partial degradation of the conjugated conduction path along the length of the nanotube, as demonstrated by a reduction in the electronic density of states near the Fermi level. Further, our results demonstrate that the reaction is controllable at low ozone exposures.

**Experiments**

Nanotubes are grown on silicon wafers with a 500 nm $SiO_2$ capping layer at 900 $^o$C via chemical vapor deposition (CVD) with methane feedstock, using a recipe that minimizes the amount of amorphous carbon deposited. Electron-beam lithography is used to control the position of the catalyst in order to create small mats of nanotubes as well as isolated individual nanotubes. For individual single-walled CVD grown nanotubes, atomic force microscopy is used to confirm that they are not bundles, and the catalyst and end caps are buried beneath the electrical contacts. Electron beam lithography and metal evaporation are used to form titanium/gold contacts and electrical transport measurements are performed using two probe and four probe ac lock-in techniques. Four probe measurements are accompanied by simultaneous two-probe measurements, as shown schematically in the optical image in Fig. 1b. XPS and UPS studies of ozone exposure use nanotubes that have been grown with laser ablation, and purchased from Tubes@Rice. The nanotubes are suspended in dichloroethane via sonication, drip-deposited on oxidized silicon wafers, and the solvent is allowed to evaporate. In both cases, the nanotubes are used without further purification. We choose CVD grown nanotubes for the transport measurements to avoid damage or contamination by solvents during post-growth processing. Since it is difficult to grow a large film of nanotubes by CVD without generating a significant concentration of multi-walled nanotubes, we use laser ablation grown nanotubes for the XPS and UPS. Though nanotubes fabricated with different growth techniques are used, previous experiments have shown no significant difference during chemical functionalization.[44] Figures 1a and 2 show scanning electron microscope images of completed nanotube mat devices that have been grown by CVD and laser ablation, respectively. Both laser ablation and CVD give a random assortment of semiconducting and metallic nanotubes, with average diameters in the range 1.1 – 1.7 nm.[45,46] The use of laser ablation grown and CVD nanotubes is important because it allows us to determine the effect of ozone on the nanotubes to the exclusion of the effect on the impurities present. The CVD growth is tuned to practically eliminate the concentration of amorphous carbon, especially in the case of the individual isolated nanotubes. In addition, though metal catalysts remain in small concentrations in the mats of nanotubes, the catalysts can be fully passivated by the gold contacts on the individual nanotubes. The concentration of metal catalysts is beneath the resolution of the XPS spectrometer and should not affect the oxidation of the nanotubes since metal-assisted oxidation has been shown to occur only at elevated temperatures.[47] Though impurities can affect the detailed oxidation rate, they do not affect the qualitative effects of ozone on the nanotubes.

Raman spectroscopy is performed in a backscattering geometry with a 514.5 nm (2.41 eV) $Ar^+$ laser. The laser is focused to a ~5 μm spot using a 100X microscope objective. The power at the sample is kept below 5 kW/cm$^2$ in order to avoid changes in the Raman spectra due to sample heating.[48-50] Raman spectra are acquired using a Kaiser Holospec spectrometer with a liquid nitrogen cooled CCD detector. Since Raman spectroscopy of nanotubes is a resonantly enhanced process,[51] the measurements are predominately sensitive to nanotubes that are resonant with the 2.41 eV photon energy. Given our diameter distribution, this photon energy will only be resonant with semiconducting nanotubes with d ~ 1.5 nm, and metallic nanotubes with d ~ 1 nm.[51] Raman will specifically not be sensitive to small or large diameter semiconducting nanotubes or large diameter metallic nanotubes. A typical Raman survey spectrum is shown in Fig. 3 and includes all of the first order carbon nanotube peaks, namely the radial breathing mode (RBM) around 200 cm$^{-1}$, the disorder induced D-band at ~1300 cm$^{-1}$, and the tangential G-band around 1590 cm$^{-1}$.[51] Also present are a sharp peak at 520 cm$^{-1}$ and a broad peak around 1000 cm$^{-1}$, which are due to the silicon/silicon dioxide substrate. The substrate peaks are used as an internal standard to confirm that the sample position has not changed during the measurement and for energy calibration.

Ozone is generated in room air using a low pressure mercury lamp with intensity at 185 nm and 254 nm. A ~1 mW/cm$^2$ lamp is used for XPS and UPS measurements of bulk laser ablation grown nanotubes, and a ~10 μW/cm$^2$ lamp is used for CVD nanotube mats in order to decrease the exposure rate for the Raman spectroscopy and transport measurements. Photoelectron spectroscopies are performed in ultrahigh vacuum (P < 5x10$^{-10}$ torr) at resolutions of 0.1-0.2 eV (XPS) and 0.05-0.1 eV (UPS) using monochromatic Al$_{K\alpha}$ X-rays ($h\nu$ = 1486.6 eV) for XPS and UV light from a He I discharge source ($h\nu$ = 21.2 eV) for UPS. During UPS measurements the samples were biased -1 to -2V to compensate for differences in sample and analyzer vacuum levels. All UPS binding energies have been corrected for the applied bias and to the Fermi level of the sample by measuring the Fermi edge of a tantalum clip in direct contact with the sample.

**Results and Discussion**

In order to understand the effect of ozone on the transport properties of nanotubes, an individual single-walled carbon nanotube was exposed to ozone using the low intensity UV lamp, while the two probe resistance, shown in Fig. 4, was measured. There is a significant increase in the resistance during UV/ozone exposure, from 150 kΩ to 180 kΩ, followed by a modest recovery to ~ 175 kΩ after the UV lamp is turned off. The small decrease in the resistance after the UV is turned off can be attributed to readsorption of photo-desorbed oxygen, though we do not observe the complete recovery that has been reported previously.[43] In addition, since the ends of the nanotube have been buried under the electrode metal and are not part of the conduction path, any changes to the nanotube must occur on the sidewall of the nanotube. The permanent change in the nanotube device resistance can be due to modification of the conducting path in the single nanotube, either by doping or defect creation,[7] or by modifications to the Schottky barriers between the nanotube and metal electrodes.

In order to eliminate the effect of the contact resistance, we perform simultaneous four probe and two probe resistance measurements on a mat of nanotubes during ozone exposure, as shown in Fig. 5. Changes in the two probe resistance are due to modifications to the entire device, possibly including the contacts; changes in the four probe resistance are only due to changes in the nanotube mat. Exposure to ozone increases both the two probe and four probe resistances by approximately 10%. The two probe resistance increases from 56.5 kΩ to 59 kΩ, while the four probe resistance increases from 9.4 kΩ to 10.1 kΩ. Since both the two probe and four probe resistances change proportionately, this indicates that the changes in resistance are dominated by effects in the nanotubes themselves and not the metal-nanotube contacts. Further, since we see the same effect on an individual nanotube, inter-nanotube junctions do not play a major role either. As we support below, we propose that the ozone oxidizes the nanotube and disrupts the π-bonded conduction path on the sidewalls of the nanotube, either at pre-existing or newly created defect sites.[24,38]

While electronic measurements allow us to determine where the ozone is modifying the nanotube and how the conductivity is changed, they do not provide information about the specific chemistry involved. For this purpose we use Raman spectroscopy to monitor the progression of the oxidation (Fig. 6). Before oxidation, the Raman spectrum of the nanotube mat shows three RBM peaks (191, 205, and 219 cm$^{-1}$) which are associated with nanotubes that are ~ 1.3, 1.2, and 1.1 nm in diameter, respectively, as well as three distinct G-band peaks.[52] Except for a small initial increase in the intensity of the D-band at 1320 cm$^{-1}$, which could be due to the creation of new defect sites on the nanotube sidewalls, the intensity of all nanotube Raman peaks decrease during oxidation (Fig. 6a). After 6 minutes of exposure, the intensity of the RBM is partially depleted, and the G-band is highly attenuated. There is also a slow shift in the position of the primary G-band peak towards higher binding energy, from 1588 cm$^{-1}$ to 1596 cm$^{-1}$. Even though the absolute intensity of the D-band decreases slightly, the ratio of the D- to G-bands, a commonly used qualitative measure of defect density,[53,54] increases linearly with time as seen in Fig. 6b. This increase is evidence for the disruption of the sidewall due to oxidation. At 7 minutes, the intensity of the RBMs is abruptly extinguished, and the shape of the G-band dramatically changes character. The lower energy modes of the G-band abruptly disappear, and the lineshape acquires an asymmetric Breit-Wigner-Fano profile characteristic of nanotube ensembles and amorphous carbon.[51] At this point, based on the extinction of the breathing modes, we conclude that the nanotubes in resonance with the laser have been entirely destroyed by the ozone. In contrast, the ratio of the D- to G-band continues to increase steadily, even after the destruction of the resonant nanotubes. The continued increase of this ratio likely has two sources. First, previous studies have shown that small diameter nanotubes are oxidized faster than large diameter nanotubes.[29,30,55,56] Since our Raman measurements are predominately sensitive to small diameter nanotubes (which are resonant with our excitation laser), larger diameter nonresonant nanotubes likely remain and show G-band Raman scattering as observed. Second, the G-band peak shift at long exposure times suggests the formation (or uncovering) of amorphous carbon.[57,58] The continued increase in the D- to G-band ratio could be due to the further oxidation of the amorphous carbon. Imaging of the devices after ozone exposure (not shown) confirms that nanotubes remain on the surface after the extinction of the nanotubes that are resonant with the Raman laser. Encouragingly, the linear increase of the D- to G-band ratio, coupled with the transport data suggest that short exposures to UV-generated ozone can controllably oxidize nanotubes.

Though Raman spectroscopy suggests the disruption of the nanotube sidewalls by oxidation, it does not provide information about the chemical species resulting from the oxidation, nor does it directly address the mechanism underlying the increase in electrical resistance upon oxidation. To understand these issues, mats of nanotubes grown via laser ablation are studied using photoelectron spectroscopies to directly determine the oxidized species and to probe the valence band electronic structure. Prior to UV/ozone exposure, the carbon 1s core level, (Fig. 7a) is dominated by a single peak at 285 eV, which is characteristic of graphitic carbon. There is also a broad peak near 291 eV. This peak is associated with a 'shake-up' process in which electron ejection also excites a $\pi-\pi^*$ transition, leading to a peak at an apparent binding energy several eV higher than the main C 1s peaks. After the nanotubes are oxidized for one hour using the 1 mW/cm$^2$ UV lamp, XPS measurements of the carbon 1s core level (Fig. 7b) confirm the increased presence of oxidized carbon species, specifically ethers or epoxides, and carbonyls,[36,38] with binding energies around 287 eV and 289 eV, respectively. In addition, the intensity of the $\pi-\pi^*$ transition is diminished, suggesting the removal of the aromatic character of the carbon nanotubes.[59] SEM investigations indicate that extensive portions of the nanotube mat are oxidized, including both the residual amorphous carbon as well as the nanotube structures (supplementary material). XPS spectra of both the as-deposited and oxidized nanotubes show no indication of metal catalysts. Since previous studies have shown that metal-assisted oxidation removes the carbon coating and exposes the catalyst,[47] the absence of catalyst core levels indicates that metal-assisted oxidation is not occurring during the UV/ozone exposure. The presence of carbonyls and epoxides could arise from the decomposition of the short-lived ozonides predicted by Criegee's mechanism, or from direct oxidation by the UV-excited singlet oxygen. It is also interesting to note that the ratio of the oxidized carbon to that of the elemental carbon in XPS increases by a factor of seven. This ratio is close to the measured increase in the ratio of D- to G-bands from Raman measurements (not shown), suggesting that the increased defect density seen in Raman spectroscopy is due to the oxidation of the nanotube wall.

Since conductivity is associated with states at the Fermi level, we use UPS to study the valence band structure of the oxidized nanotube films. The valence band structure of the as-deposited and oxidized nanotube films used for XPS are shown in Fig. 8, where the spectra have been normalized such that zero binding energy is equivalent to the sample Fermi level. After exposure to ozone, there is a reduction in the density of electronic states (DOS) near the Fermi level as seen in the inset to Fig. 8. This is due to the disruption of the π-conjugation,[60] correlating with the Raman and XPS data. Further, since conductivity depends on the number of electrons near the Fermi level, the loss of the π-states leads to the observed increase in electrical resistance. Second, there is a 1 eV shift of the high binding energy cutoff towards lower binding energy. In the photoelectric effect, the work function is related to the high binding energy cutoff, $\Phi = h\nu - E_{cutoff}$, so the 1 eV decrease in the high binding energy cutoff means a 1 eV increase in the work function. This increase can be attributed to both the reduction in the π-conjugation, as well as the increase in the surface dipole due to the oxygen containing functional groups.[61] Work function increases and DOS decreases have been observed previously for multi-walled nanotubes that were acid, air, or oxygen plasma oxidized.[61] There are, however, theoretical claims that ozone physisorption or chemisorption onto single-walled nanotubes will lead to an *increase* in DOS near the Fermi level, and a *decrease* in the electrical resistance.[4] An experiment on *multi*-walled nanotubes appears to support this result.[62] Based on the data we present here, we conclude that the dominant overall effect of ozone on the *exposed* wall of a nanotube is a reduction of the π-conjugated states due to the addition of oxidized carbon species. It thus appears that the effects of ozone on nanotubes include the creation of additional chemical groups not yet addressed by theory.[4]

**Summary**

Raman spectroscopy and transport measurements have been used to study the ozone oxidation of carbon nanotubes. The electrical resistance of the nanotube increases upon exposure to ozone and is irreversible. Comparisons between nanotube mats and individual nanotubes indicates that the resistance change is due to sidewall oxidation and disruption of the conduction network on the individual nanotube level, rather than being caused by end cap oxidation, destruction of the inter-tube contacts, or photo-desorption induced changes in the metal-nanotube Schottky barrier. Raman spectroscopy, XPS, and UPS confirm that ozone extensively oxidizes the nanotube and causes a significant disruption of the conjugated $\pi$-bonding on the nanotube sidewall. UPS measurements show a 1 eV increase in the work function of the oxidized nanotubes, accompanied by a loss of electronic states near the Fermi level, leading to the increased electrical resistance. Finally, transport and Raman data suggest that exposure of nanotubes to ozone may lead to controllable oxidations in short times and at room temperature.

**Acknowledgements**: The authors acknowledge financial support from the NSF MRSEC program under Grant No. DMR-0079983, the NSF CAREER program under Grant No. DMR-0094063, the NSF NSEC program under Grant No. DMR-0425880, and NSF DMR-0210806.

**Supporting Information Available**: SEM images of single-walled nanotube mats before and after UV/ozone exposure. This material is available free of charge via the Internet at http://pubs.acs.org.

**Figures:**

1. (a) SEM image of a CVD grown nanotube mat device with metal contacts. (b) Optical image of a completed CVD device, with schematic circuit diagram for four probe transport measurements. The purple background is the $SiO_2$ substrate, the yellow regions are the Ti/Au contacts, and the olive region is the location of the nanotube mat.

2. SEM image of a mat of laser-ablation grown nanotubes used for Raman and photoelectron spectroscopies.

3. Raman spectrum of a CVD grown mat of nanotubes. The nanotube peaks consist of radial breathing modes (RBM) around 200 cm$^{-1}$, a $sp^3$-like disorder band (D) at ~1300 cm$^{-1}$, and $sp^2$-like tangential band (G) at 1590 cm$^{-1}$. Peaks labeled with an asterisk (*) come from the $Si/SiO_2$ substrate.

4. Transport data for an individual single-walled nanotube during ozone oxidation.

5. Two probe (a) and four probe (b) resistance of a CVD-grown nanotube mat during ozone exposure. The similarity between the two probe and four probe resistance changes indicates that ozone primarily affects the conduction in the nanotube mat and not at the nanotube-metal contact.

6. (a) Raman spectra of a mat of CVD grown nanotubes from Fig. 1a, during ozone oxidation. During ozone exposure, the intensity of all nanotube bands decreases. At 7 minutes, the intensity of the RBM is abruptly extinguished, accompanied by a change in G-band lineshape, indicating that the nanotubes resonant with the 2.41 eV photon energy have been removed. (b) Ratio of the area of the D-band to the G-band area.

7. XPS spectra of the nanotube mat in Fig. 2 before and after a 1 hour exposure to ozone generated by a high intensity UV lamp in air. After ozone exposure, there is a large increase in the intensity of the ether and carbonyl core levels, and a decrease in the intensity of the $\pi-\pi^*$ shake-up peak (see text).

8. UPS valence band structure of the nanotube mat of Fig. 7, before (solid line) and after ozone exposure (dashed line). The intensity of the conjugated $\pi$-bonds is decreased after ozone exposure, leading to decreased density of states near the Fermi level (at 0 eV). The increased work function is apparent from the shift in the high-binding energy cut-off. Inset: expanded view near the Fermi level highlighting the disruption of the $\pi$-states.

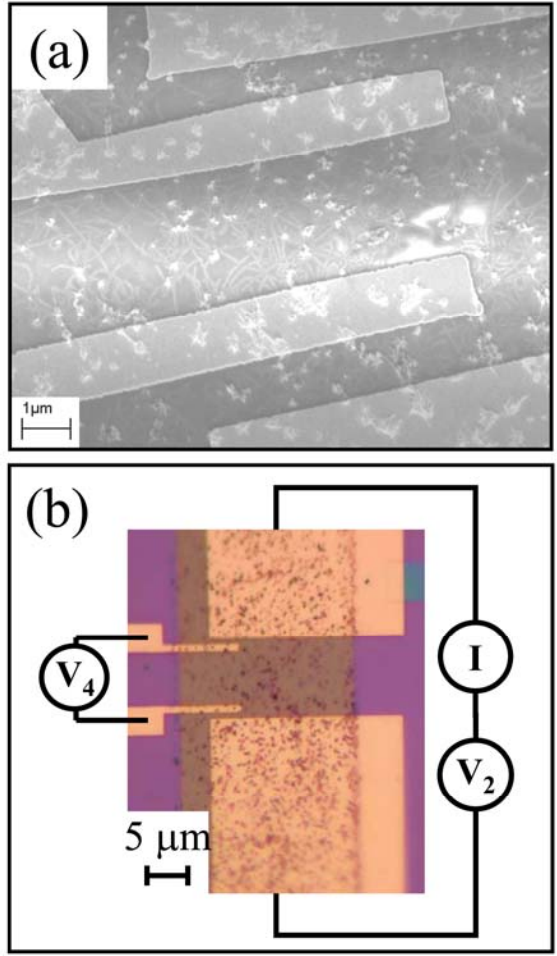

Figure 1: (a) SEM image of a CVD grown nanotube mat device with metal contacts. (b) Optical image of a completed CVD device, with schematic circuit diagram for four probe transport measurements. The purple background is the $SiO_2$ substrate, the yellow regions are the Ti/Au contacts, and the olive region is the location of the nanotube mat.

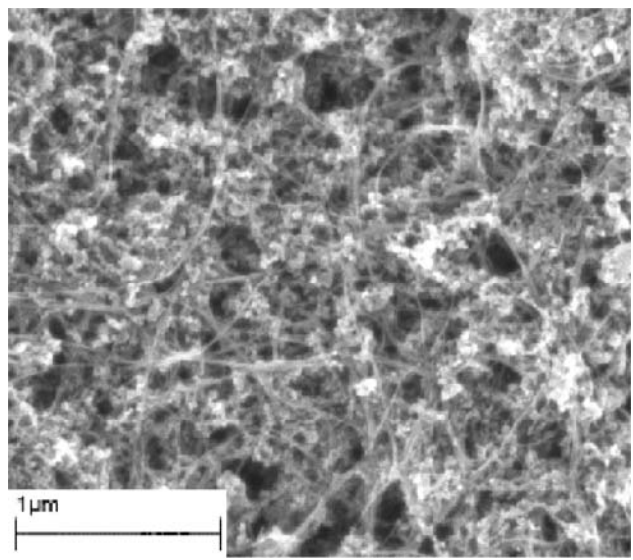

Figure 2: SEM image of a mat of laser-ablation grown nanotubes used for Raman and photoelectron spectroscopies.

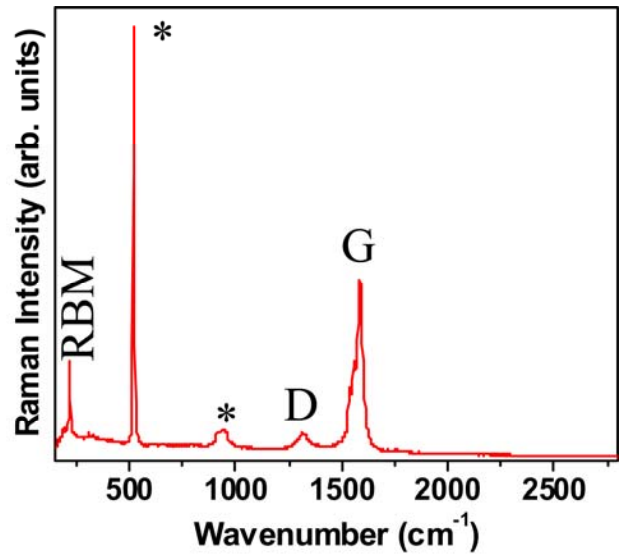

Figure 3. Raman spectrum of a CVD grown mat of nanotubes. The nanotube peaks consist of radial breathing modes (RBM) around 200 cm$^{-1}$, a *sp$^3$*-like disorder band (D) at ~1300 cm$^{-1}$, and *sp$^2$*-like tangential band (G) at 1590 cm$^{-1}$. Peaks labeled with an asterisk (*) come from the Si/SiO$_2$ substrate.

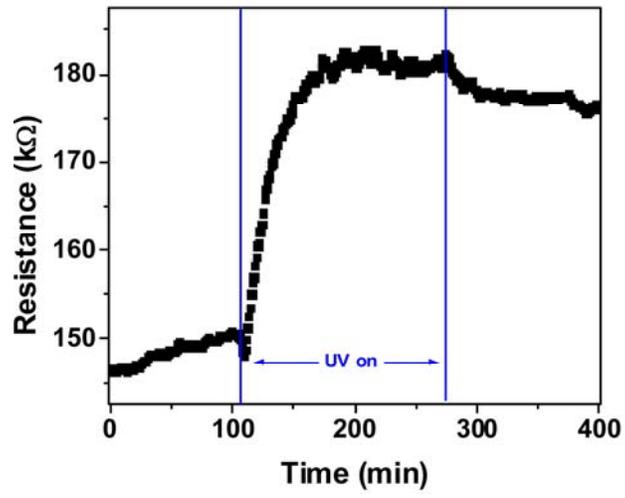

Figure 4: Transport data for an individual single-walled nanotube during ozone oxidation.

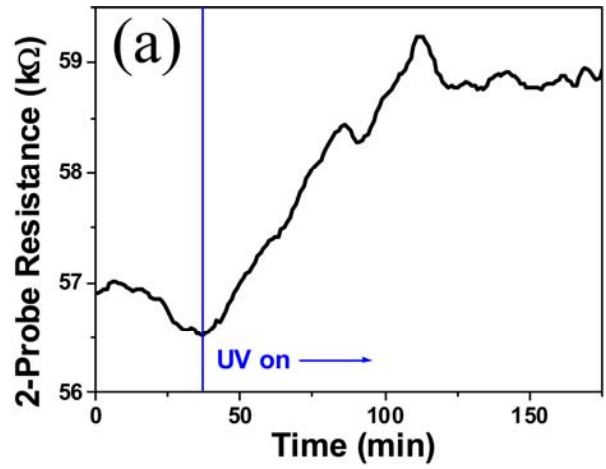

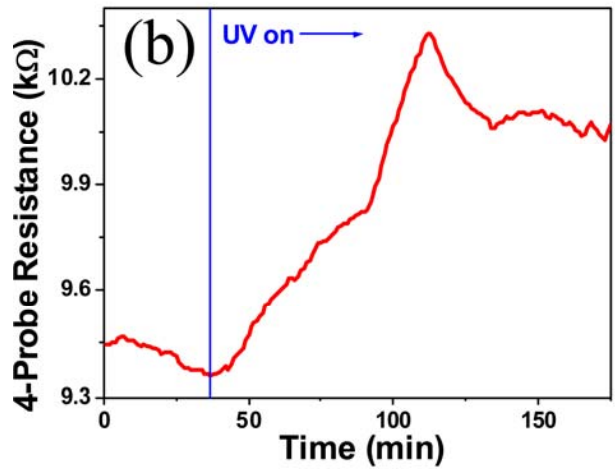

Figure 5: Two probe (a) and four probe (b) resistance of a CVD-grown nanotube mat during ozone exposure. The similarity between the two probe and four probe resistance changes indicates that ozone primarily affects the conduction in the nanotube mat and not at the nanotube-metal contact.

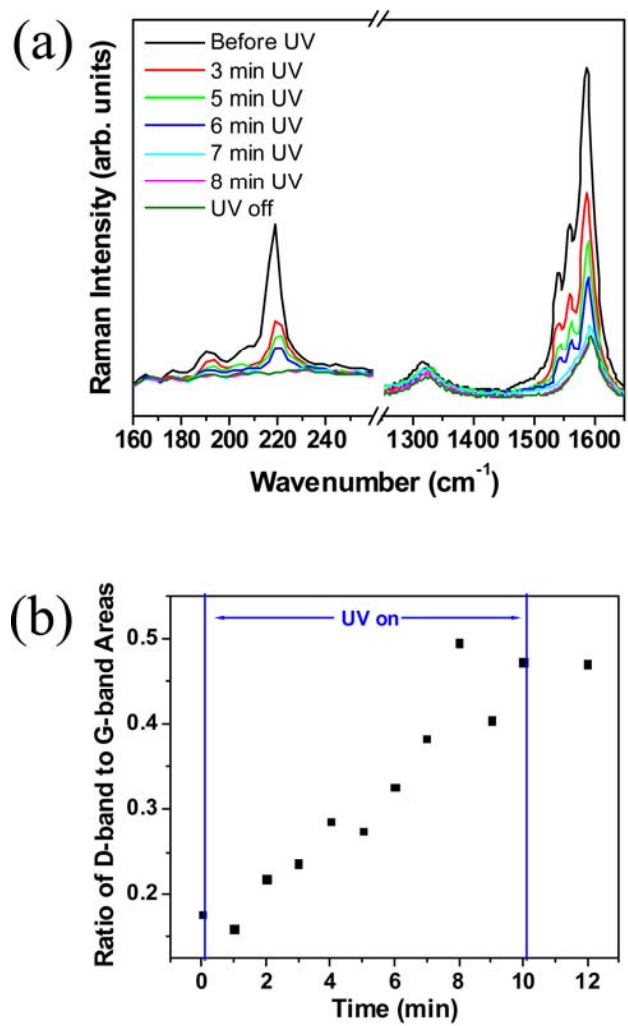

Figure 6: (a) Raman spectra of a mat of CVD grown nanotubes from Fig. 1a, during ozone oxidation. During ozone exposure, the intensity of all nanotube bands decreases. At 7 minutes, the intensity of the RBM is abruptly extinguished, accompanied by a change in G-band lineshape, indicating that the nanotubes resonant with the 2.41 eV photon energy have been removed. (b) Ratio of the area of the D-band to the G-band area.

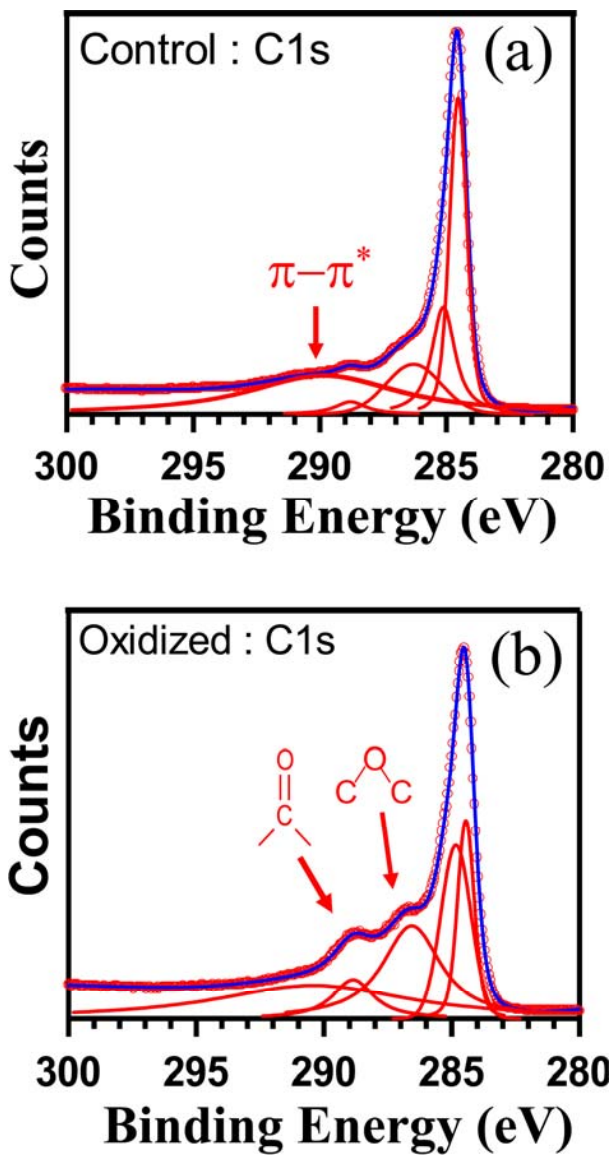

Figure 7: XPS spectra of the nanotube mat in Fig. 2 before and after a 1 hour exposure to ozone generated by a high intensity UV lamp in air. After ozone exposure, there is a large increase in the intensity of the ether and carbonyl core levels, and a decrease in the intensity of the π−π* shake-up peak (see text).

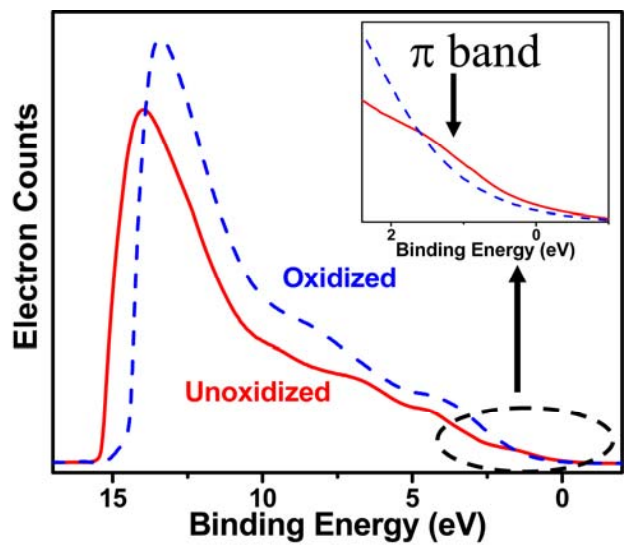

Figure 8. UPS valence band structure of the nanotube mat of Fig. 7, before and after ozone exposure. The intensity of the conjugated π-bonds is decreased after ozone exposure, leading to decreased density of states near the Fermi level (at 0 eV). The increased work function is apparent from the shift in the high-binding energy cut-off. Inset: expanded view near the Fermi level highlighting the disruption of the π-states.

**Supplementary Information:**

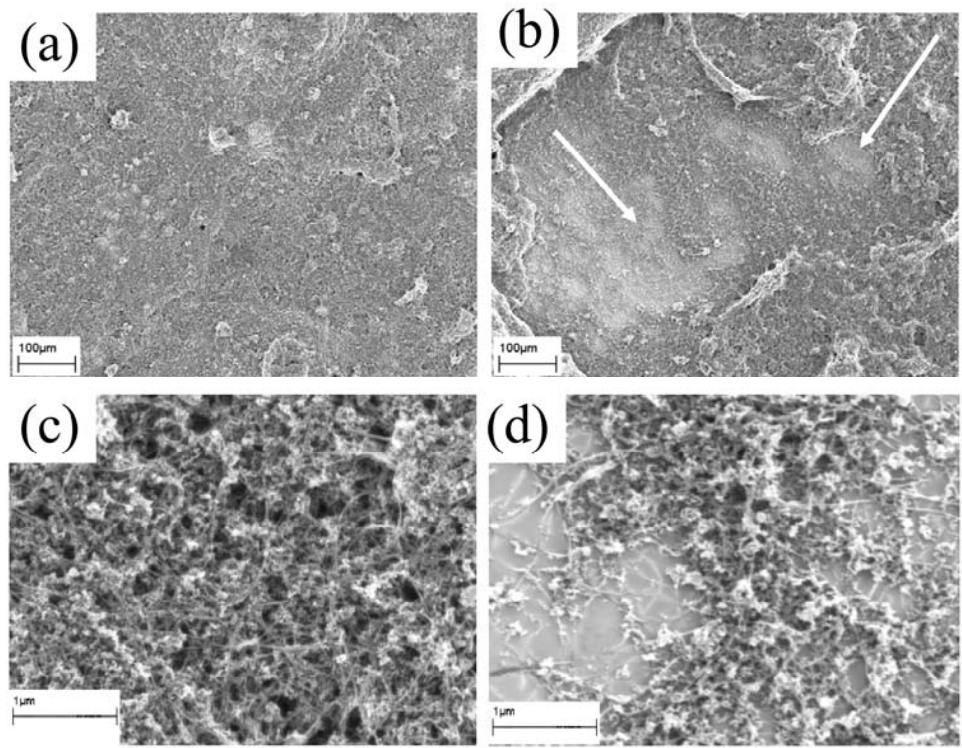

Figure S1: SEM micrographs showing the large scale removal of carbon from a mat of single walled nanotubes. (a) and (c) are low and high magnification images of the initial mat of single-walled nanotubes before exposure to UV/ozone. The mat is fairly uniform and opaque over large length scales. After UV/ozone exposure, low magnification imaging, (b), shows extensive etching of the nanotube mat, exposing the substrate in some regions (arrows). High magnification imaging, (d), shows that both amorphous carbon and nanotubes are etched.